\documentclass[graybox]{svmult}

\usepackage{amsmath}
\usepackage{mathptmx}       
\usepackage{helvet}         
\usepackage{courier}        
\usepackage{makeidx}         
\usepackage{graphicx}        
\usepackage{multicol}        

\makeindex             

\begin{document}

\title*{A dynamical view of different solution paradigms in
two-person symmetric games: Nash vs. co-action equilibria}
\titlerunning{A dynamical view of Nash and co-action equilibria in games}
\author{V. Sasidevan and Sitabhra Sinha}
\institute{V. Sasidevan \at The Institute of Mathematical Sciences,
CIT Campus, Taramani, Chennai 600113, India.
\email{sasidevan@imsc.res.in}
\and Sitabhra Sinha \at The Institute of Mathematical Sciences, 
CIT Campus, Taramani, Chennai 600113, India.
\email{sitabhra@imsc.res.in}
}
%
%
\maketitle

\abstract*{The study of games and their equilibria is central to developing insights for understanding many socio-economic phenomena. Here we present a dynamical systems view of the equilibria of two-person, payoff-symmetric games. In particular, using this perspective, we discuss the differences between two solution concepts for such games - namely, those of Nash equilibrium and co-action equilibrium. For the Nash equilibrium, we show that the dynamical view can provide an equilibrium refinement, selecting one equilibrium among several possibilities, thereby solving the issue of multiple equilibria that appear in some games. We illustrate in detail this dynamical perspective by considering three well known 2-person games namely the Prisoner's Dilemma, game of Chicken and the Stag-Hunt. We find that in all of these cases, co-action equilibria tends to correspond to `nicer' strategies than those corresponding to Nash equilibria.}

\abstract{The study of games and their equilibria is central to developing insights for understanding many socio-economic phenomena. Here we present a dynamical systems view of the equilibria of two-person, payoff-symmetric games. In particular, using this perspective, we discuss the differences between two solution concepts for such games - namely, those of Nash equilibrium and co-action equilibrium. For the Nash equilibrium, we show that the dynamical view can provide an equilibrium refinement, selecting one equilibrium among several possibilities, thereby solving the issue of multiple equilibria that appear in some games. We illustrate in detail this dynamical perspective by considering three well known 2-person games namely the Prisoner's Dilemma, game of Chicken and the Stag-Hunt. We find that in all of these cases, co-action equilibria tends to correspond to `nicer' strategies than those corresponding to
Nash equilibria.}

\section{Introduction}
\label{sec:1} 
Games represent strategic interactions between entities generally referred to as agents. Here, the term \textquotedblleft agents\textquotedblright\; could refer to a variety of entities, ranging from human beings or animals to computer programs or robots. In games, each agent receives a payoff depending upon the strategy choice made by  all agents including herself. Thus, an agent who wants to optimize her payoff should consider not only the payoff structure of the game, but also the decision making processes of other agents. The choice of strategy by each agent in such an interaction leads to a collective outcome that may or may not be globally optimal. In  this context, it is imperative to understand how two agents facing a game situation, who have to make a strategic decision, will go about doing it, since the strategic interaction between agents is the  basis of the  collective behavior  in a system  comprising such agents.  Financial markets, for example, may be viewed as the collective outcome  of 
strategic interactions between a large number of people participating in it. 
Another example is that of evolution, where one may view evolution by natural selection  as a result of the interaction between competing genes. Cooperation and conflict is at the heart of such systems and forms the subject matter of the study of games. In games, in general, each agent should have a behavior model of other agents  so that she has a way to describe the  decision making process of other agents. In this regard, standard game theory makes several assumptions about the agent's behavior. It assumes that agents are fully rational and would like to optimize their payoff and they are perfect in execution of their strategies  (see for e.g \cite{hargreaves} for a detailed discussion).  While  the applicability of  these assumptions in any particular situation is open to criticism, they form an important benchmark for optimal behavior. In fact, these assumptions form an important part of  modern economic theory in which the participating agents are often assumed to be fully rational. 

The simplest of games consists of the strategic interaction between two agents in a single play of the game. In fact, 2-person games like Prisoners Dilemma, Stag-Hunt etc., describe very general socio-economic scenarios, towards the analysis of which considerable effort has been devoted. A key concept in the study of games is that of an \textquotedblleft equilibrium\textquotedblright\;. It refers to a state of affairs where each agent has decided her strategy  for the  game at hand. How the agents pick their equilibrium strategy is given by a solution concept. A solution concept thus is a formal rule for predicting how a game will be played between agents and employs certain assumptions regarding agent's behavior. An important solution concept for non-cooperative games is that of Nash equilibrium. Informally, it is a state where after every agent has selected their `Nash' strategies, none of the agents can improve their payoff by unilaterally deviating from it. It is to be noted that a game may have more 
than one Nash equilibrium.

In this article, we show that the equilibria of a game may be viewed as the \textquotedblleft fixed-point\textquotedblright\; equilibria of a dynamical system. In particular, we present a dynamical view of the equilibria obtained
by two different solution concepts, viz., Nash~\cite{osborne} and co-action~\cite{sasidevan}, the latter being a concept that makes use of the symmetry between the agents for payoff-symmetric games. The vector
flow diagrams on the strategy space that is generated using the dynamics approach makes the differences
between the equilibria obtained in the two solution concepts visually apparent. For the Nash equilibrium, we argue that a dynamical perspective may be regarded as an equilibrium refinement selecting one equilibrium out of several possible ones, thus solving the multiplicity issue. We illustrate these points by considering  three well known examples of 2-person games, namely the Prisoners Dilemma, Game of Chicken and Stag-Hunt. 
 
\section{A dynamical framework for analyzing 2-person games}
\label{sec:2}
\begin{figure}[tbp]
\sidecaption
\includegraphics[width=.5\linewidth]{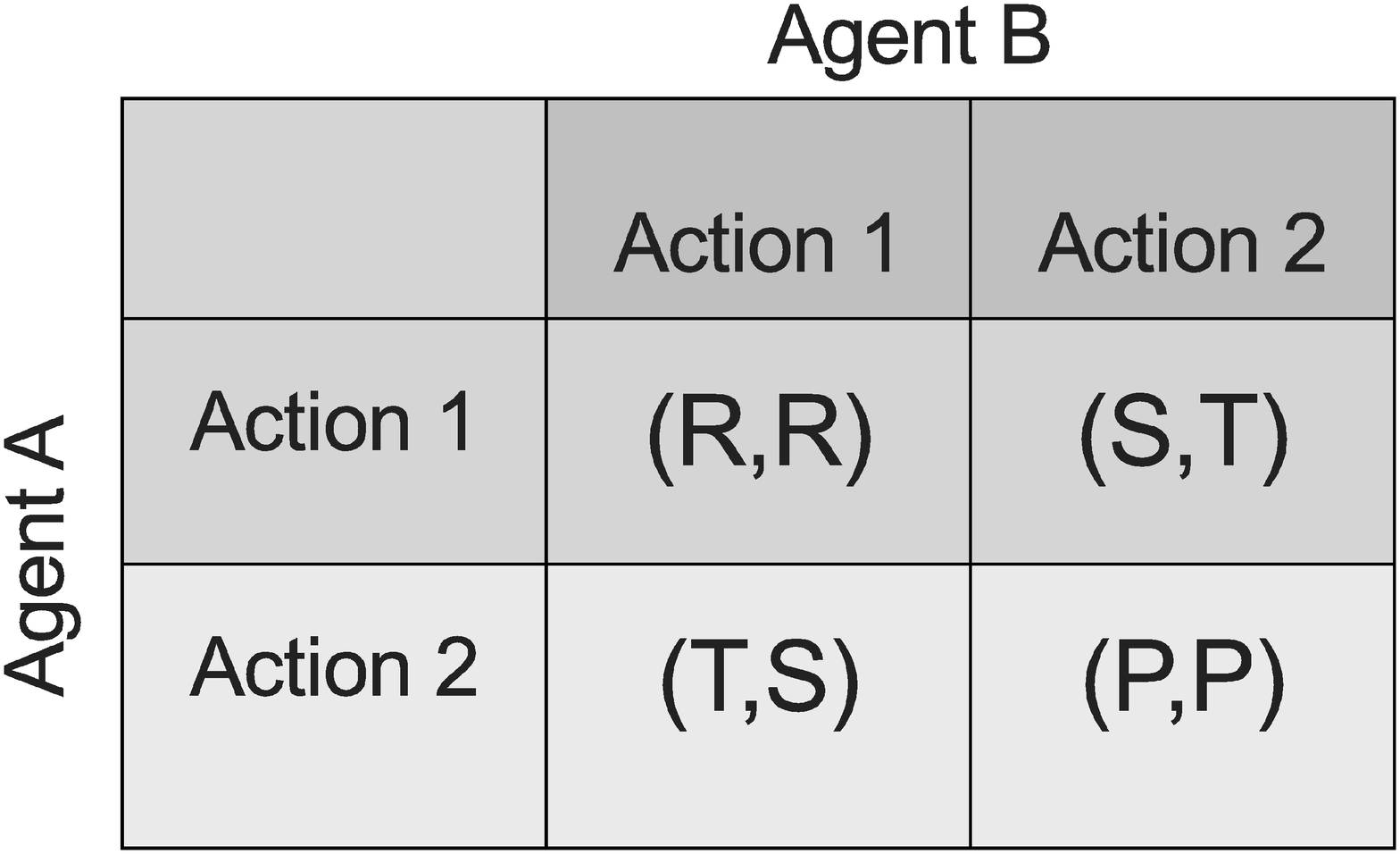}
\caption{A generic representation of the payoff matrix for a 2-person
symmetric game where each agent has two actions available to her.}
 \label{fig:0}
\end{figure}

Here we describe a dynamical perspective for analyzing games,
focusing on 2-person single-stage games in which two agents interact
only once. No communication is allowed between the agents. Furthermore, we consider the simple case where each agent
has to choose one of two possible actions (say, Action~1 and Action~2)
available to her. Each agent receives a payoff according to the pair
of choices made by them, such that the game may be represented by a
payoff matrix that specifies all  possible outcomes
(Fig.~\ref{fig:0}). We consider situations where the game is payoff
symmetric, i.e., on exchanging the identities of the players (A,B),
the payoff matrix remains unchanged. Note that most 2-person games
that are studied in the literature fulfil the above criteria. Given
the payoff matrix, an agent can have a mixed strategy, where she
chooses Action~1 with some probability $p$ and Action~2 with
probability $(1-p)$. If $p$ is either 0 or 1, it is called a pure
strategy. Given a game, represented by a matrix containing the
numerical values of $R,S,T$ and $P$ (or a hierarchical relation among
them), Nash equilibrium is defined as a state -
i.e., a set of the choices made by all the agents - where no agent can
increase her payoff by unilaterally deviating from the Nash state. A
Nash equilibrium comprising pure strategies may be found by a search
procedure, whereby each possible state is explicitly examined for the
above criterion. Note that a given game can have more than one Nash
equilibrium, possibly involving mixed strategies. In such cases, the
choice of a particular equilibrium will have to involve additional
refinement criteria, which is an important area of research in game
theory~\cite{harsanyi88}.

\begin{figure}[tbp]
\includegraphics[width=.9\linewidth]{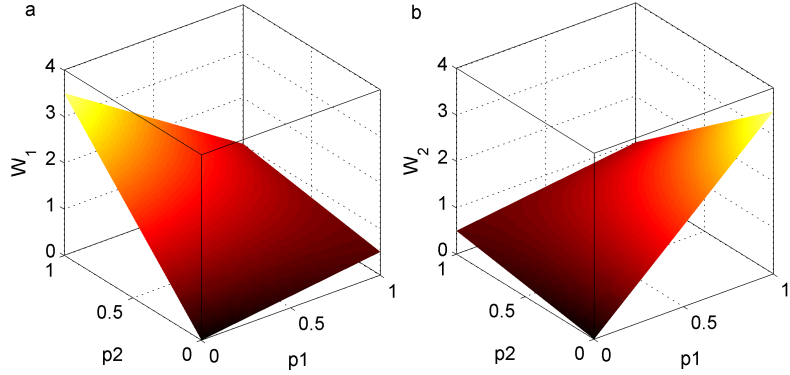}
\caption{The payoff functions $W_A$ and $W_B$ for two agents playing 
the game of Chicken, shown 
as functions of $p_1$ and $p_2$, i.e., the probability of each agent
to choose Action~1. The payoffs (in terms of the terminology
given in Fig.~\ref{fig:0}) are $T=3.5$, $R=1$, $S=0.5$ and $P=0$.}
\label{fig:1}       
\end{figure}

We now illustrate a dynamical perspective on Nash equilibria by first
defining payoff functions for all possible mixed strategies of the two
agents. Assuming that agent A (B) chooses Action~1 with probability
$p_1$ ($p_2$) and Action~2 with probability $1-p_1$ ($1- p_2$,
respectively), the expected payoffs of the agents are
\begin{align}
 W_A &= p_1 p_2 R + p_1 (1 - p_2) S + (1-p_1) p_2 T + (1-p_1)(1-p_2) P,\\
 W_B &= p_1 p_2 R + p_1 (1 - p_2) T + (1-p1) p_2 S + (1-p1) (1-p2) P.
\end{align}
As the payoffs are continuous functions of $p_1$ and $p_2$, they can
be represented as two-dimensional surfaces (Fig.~\ref{fig:1})
analogous to fitness landscapes in biology or energy landscapes in
physics. However, unlike the latter, there are two distinct surfaces
for the two agents, and each of them would like to achieve the maximum
of their respective payoff functions, a goal that may not be mutually
compatible. By contrast, the evolution of the state of a physical
system can be seen as a convergence process to a minimum of a single
function, e.g., the free energy that describes the entire system.

%
Given the payoff function surfaces we can now proceed to find the
strategy pairs ($p^*_1, p^*_2$) that correspond to a Nash equilibrium.
Note that while the Nash solution is usually not defined in terms of a
dynamical perspective, one can view ($p^*_1, p^*_2$) as an equilibrium
point for  flow dynamics in the $p_1-p_2$ plane, as described below.
The initial condition for this dynamical system can be any arbitrary
point in this plane. Each agent is then allowed to change its strategy
infinitesimally (i.e., $p_1 \rightarrow p_1 + dp_1, p_2 \rightarrow
p_2 + dp_2$) in order to improve their respective payoffs, taking
into consideration that the other agent would also be doing the same.
A sequence of such incremental changes, which will be manifested as a
flow in the $p_1-p_2$ plane would eventually converge to an equilibrium
point ($p^*_1, p^*_2$). Note that, while such a strategy would
correspond to a stable equilibrium of the flow dynamics, there may
also be unstable equilibria.

The dynamical equations governing the flow  can be derived by
considering the change in the payoffs ($dW_A, dW_B$) of the two agents
as a result of the infinitesimal change in their strategies $dp_1,
dp_2$:
\begin{align}
 \nonumber \dfrac{\partial W_A}{\partial p_1} &= p_2 (R-T) + (1-p_2) S,\\
  \dfrac{\partial W_B}{\partial p_2} &= p_1 (R-T) + (1-p_1) S.
 \label{eq:1}
\end{align}
Thus, on any point in the $p_1-p_2$ plane, the magnitude and direction
of the flow can be obtained by a vector sum of the two components
given by Eq.~\ref{eq:1}. The resulting flow diagram will describe the
trajectory in strategy space starting from any arbitrary strategy pair ($p_1, p_2$).
This will be illustrated with specific examples of 2-person games in
the next section.

As mentioned earlier, Nash equilibrium is not the only possible
solution of a  payoff symmetric game. Recently, an alternative paradigm
referred to as co-action equilibrium
for solving such games has been introduced in the specific context of
minority game~\cite{sasidevan}. Here we study this novel solution
concept in the context of generic 2-person games with symmetric payoff
from a dynamical perspective.
The key notion of co-action equilibrium is: as the two agents are
aware that they face an exactly symmetric situation, the choice made
by agent A should be identical to the choice of agent B, assuming that
they are equally rational (for a detailed discussion see
Ref.~\cite{upcoming}). Thus, in terms of the flow dynamics introduced
above, in this solution concept, each agent will take into account in
her calculation for revising her strategy that the other agent is not
only using the same strategy (i.e., $p_1 = p_2$) but will also
make exactly the same infinitesimal change, i.e., $dp_1 = dp_2$. Then
the change in the payoffs of the two agents, as a result of changing
$p_1, p_2$ (analogous to Eq.~\ref{eq:1} for Nash equilibrium) is:
\begin{align}
\nonumber 
\dfrac{\partial W_A}{\partial p_1} &= 2 p_1 R + (1- 2 p_1) (T + S),\\
\dfrac{\partial W_B}{\partial p_2} &= 2 p_2 R + (1 - 2 p_2) (T + S).
\label{eq:2}
\end{align}
Note that the above equations hold not only when $p_1 = p_2 = p$ (so
that the dynamics is confined to the diagonal line in the $p_1-p_2$
plane), but also for situations where the two agents initially start with
different
strategies ($p_1 \neq p_2$), believing however that the other agent is
using exactly the same strategy. 

The co-action solution yields results that differ remarkably
from those obtained using the concept of Nash equilibrium, some of
which will be described in the next section in the context of specific
2-person games. An important distinction is that while there could be
multiple Nash equilibria for a game, the corresponding co-action
equilibrium is unique. The dynamical perspective allows us to also 
distinguish between Nash and co-action solutions for 2-person
symmetric games in that
a stable mixed strategy equilibrium is possible for the latter unlike
in the former (Nash) where a mixed strategy
equilibrium, if it exists,
is  always unstable. 

Note that while the flow diagrams produced by the dynamical process
presented here may resemble the trajectories generated by solving
replicator equations~\cite{Hofbauer98}, the two approaches are
essentially distinct. In particular, the latter approach is based on
the concept of evolutionary stable strategies, which is an equilibrium
refinement of the Nash solution. Also, instead of being stages in the
evolutionary progression of a population,  the sequence of infinitesimal changes in strategies in the dynamical approach
 presented here, can
be interpreted as  steps in the deductive reasoning of the two
agents, at the end of which they choose the strategy corresponding to
the equilibrium they converge to. When our approach is applied to
study the Nash solution of a game, it can also be viewed as an equilibrium
refinement as, if there are multiple Nash equilibria, it allows agents
to choose a particular equilibrium depending on the arbitrarily chosen 
initial state. Thus, in an ensemble of many realizations of a game,
the fraction of cases where agents will converge to a particular
equilibrium is proportional to the size of its basin of attraction.
An unstable equilibrium (if it exists) will lie on the separatrix that
demarcates the
basins of different stable equilibria.

\section{Examples}
\label{sec:3}
We now illustrate the approach outlined in the previous section using
three well-studied 2-person symmetric games, viz., Prisoner's Dilemma,
Game of Chicken and Stag-Hunt, each of which can be connected to
real-life scenarios involving interactions between a pair of agents
who have to choose between two possible actions. 
Each of these games is defined in terms of a specific hierarchical
relationship between the payoffs $R$, $S$, $T$ and $P$ (using the
terminology of the payoff matrix in Fig.~\ref{fig:0}). Without loss of
generality, we can set $P = 0$ and $R = 1$ (thereby fixing the origin
and the scale for the payoffs), leaving only $S$ and $T$ as free
parameters.
In the following subsections, we discuss each of these games in turn,
describing the meaning of the different choices available to the agents
(viz. Action 1 and Action 2) in a particular game, and exploring the
different equilibria obtained by using Nash and co-action solution
concepts.

\subsection{Prisoner's Dilemma}
Prisoner's Dilemma (PD)~\cite{Rapoport65} can be regarded as one of
the most well-known games in the literature. It has evoked great
interest among researchers from a multitude of disciplines
ranging from social sciences and politics to biology and
physics, from the 1950s onwards and continues to do so at present
(a good place to read about historical developments in PD is the
corresponding entry in the online Stanford Encyclopedia of
Philosophy~\cite{stanford}). The game represents the strategic interaction
between two players who have to choose between
cooperation (Action 1) and defection (Action 2). The different
payoffs are interpreted as follows: $R$ is a ``reward'' for both
players cooperating, $P$ is a ``punishment'' for both players
defecting, while, in the event that one agent defects while the other
cooperates, $T$ and $S$ are the ``temptation'' received by former and
the ``sucker's payoff'' of the latter.
In PD, the hierarchical relation between the payoffs is $T>R>P>S$,
which makes achieving mutual cooperation non-trivial as each player
will benefit more by defecting (assuming that the other will
cooperate). 

\begin{figure}[tbp]
\sidecaption[t]
\includegraphics[width=.45\linewidth]{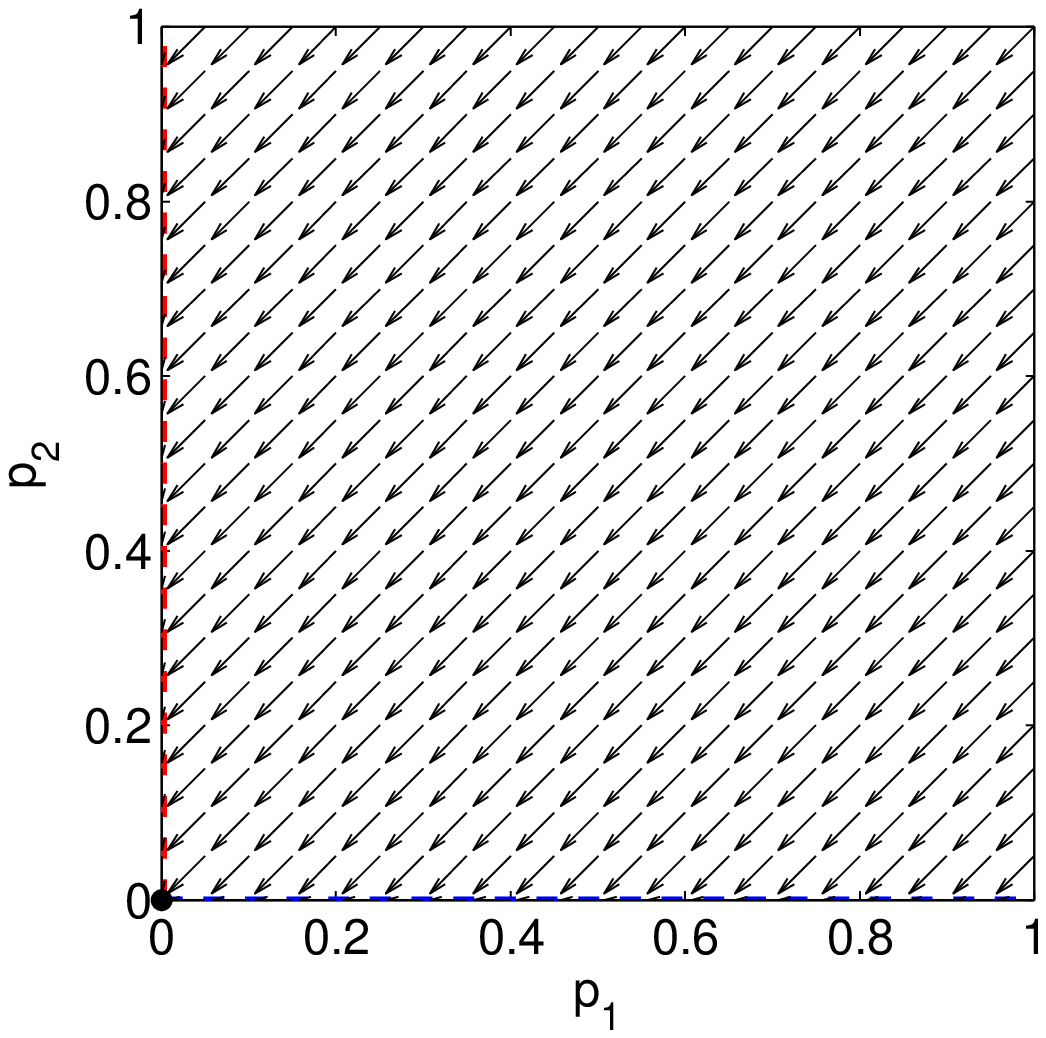}
\includegraphics[width=.45\linewidth]{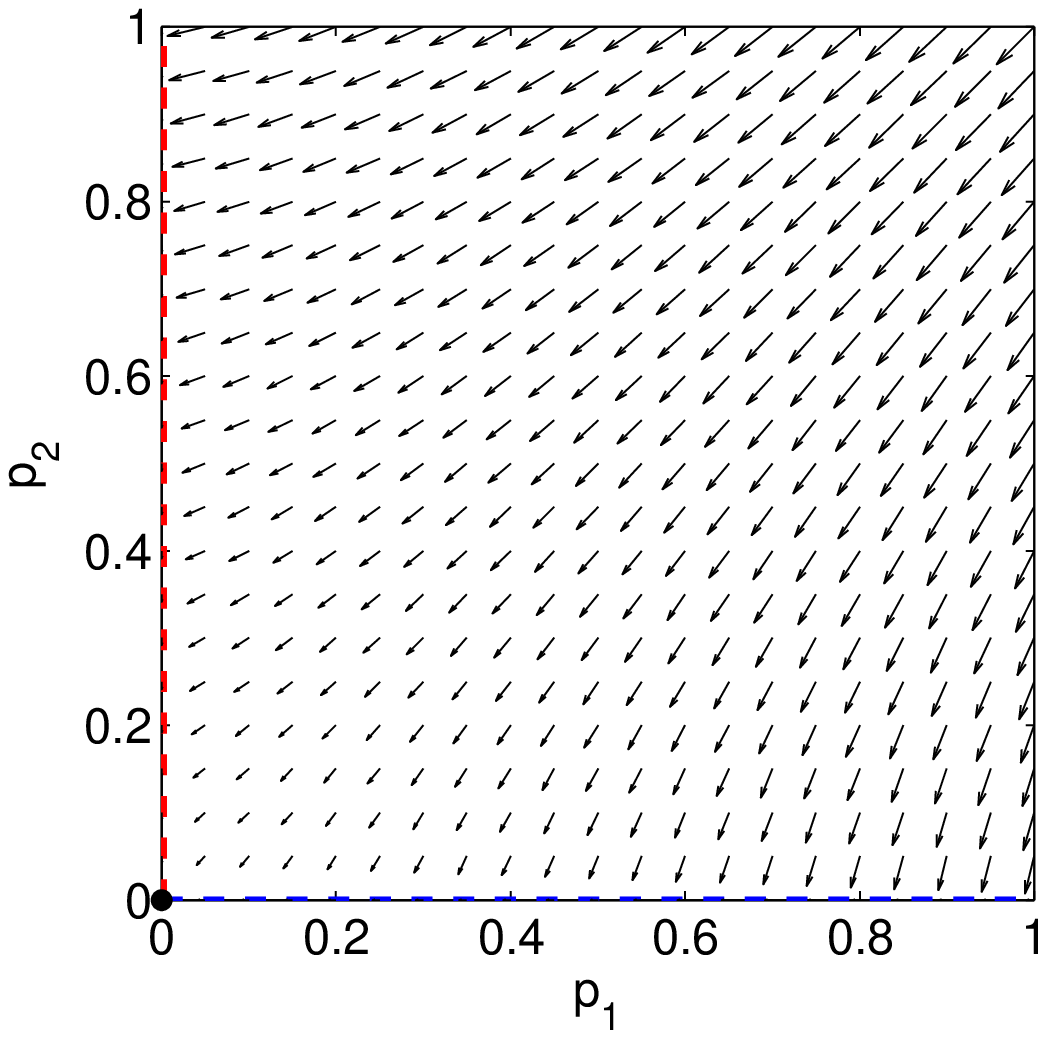}
%
%
\caption{Vector flow diagram representation of the Nash solution of the 2-person Prisoners Dilemma game for temptation payoffs (left) $T =1.5$ and (right) $T = 3.5$. The abscissae and ordinate correspond to the probabilities ($p_1$ and $p_2$) that players 1 and 2, respectively, choose to cooperate. The broken lines represent the best response (or reaction) correspondence of the players (red for player 1, blue for player 2). The intersection of the lines, represented by a filled circle, represent the single Nash equilibrium corresponding to both players defecting (i.e., $p_1=0$, $p_2=0$).
}
\label{fig:2}       
\end{figure}

It is easy to see that mutual defection is the only Nash equilibrium
for PD. As Action 1 represents cooperation, $p_1$ ($p_2$)
corresponds to the probability that agent A (B) will choose
cooperation. As discussed in the previous section, we can associate a
vector with each point in the ($p_1, p_2$) plane for the game which
describes the flow from that point. Fig.~\ref{fig:2} shows the
resulting flow diagrams obtained using the Nash solution concept for
two different values of the temptation payoff $T$ (keeping $S$ fixed
at $-0.5$). In both cases, the system converges to the pure strategy
$p_1 = p_2 = 0$ (mutual defection), which is the Nash equilibrium for
PD.

\begin{figure}[tbp]
\sidecaption[t]
\includegraphics[width=.45\linewidth]{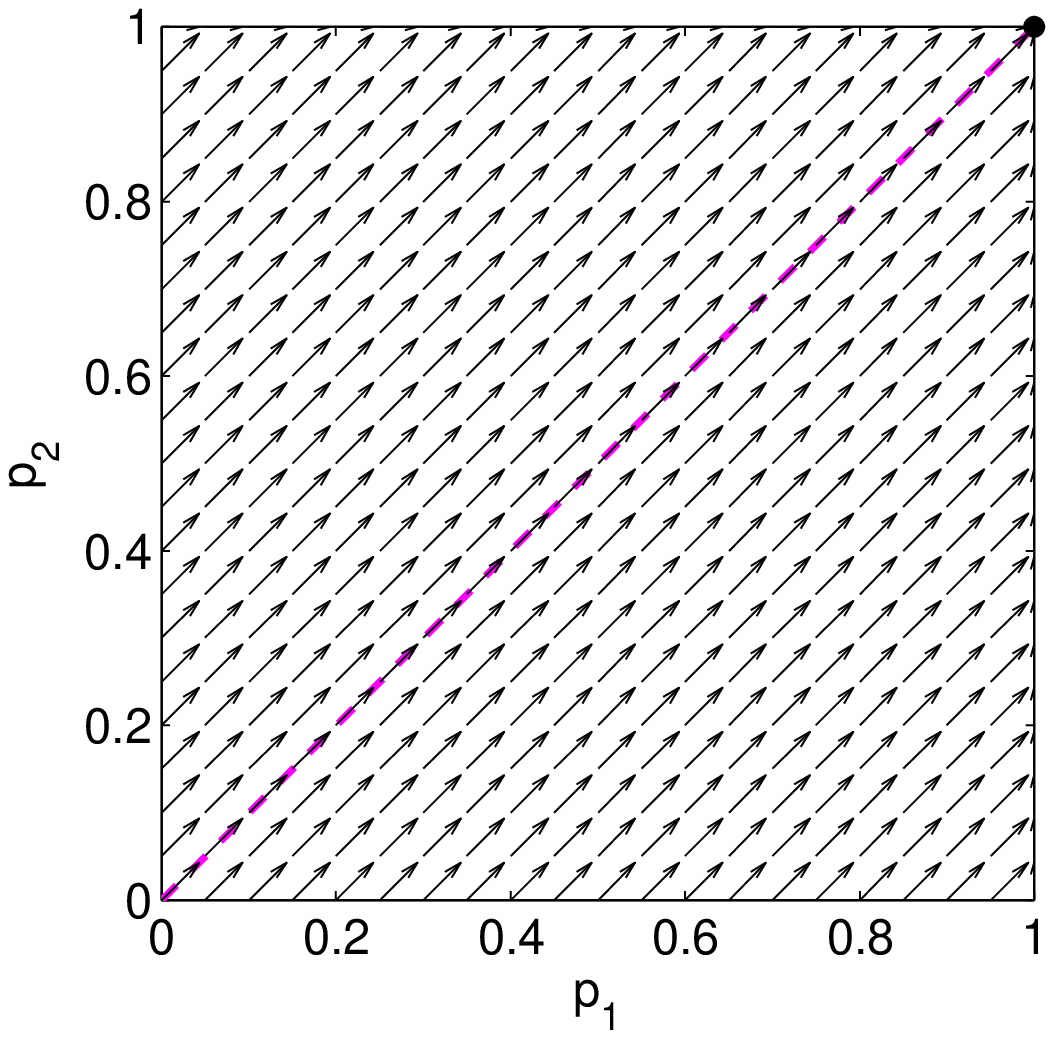}
\includegraphics[width=.45\linewidth]{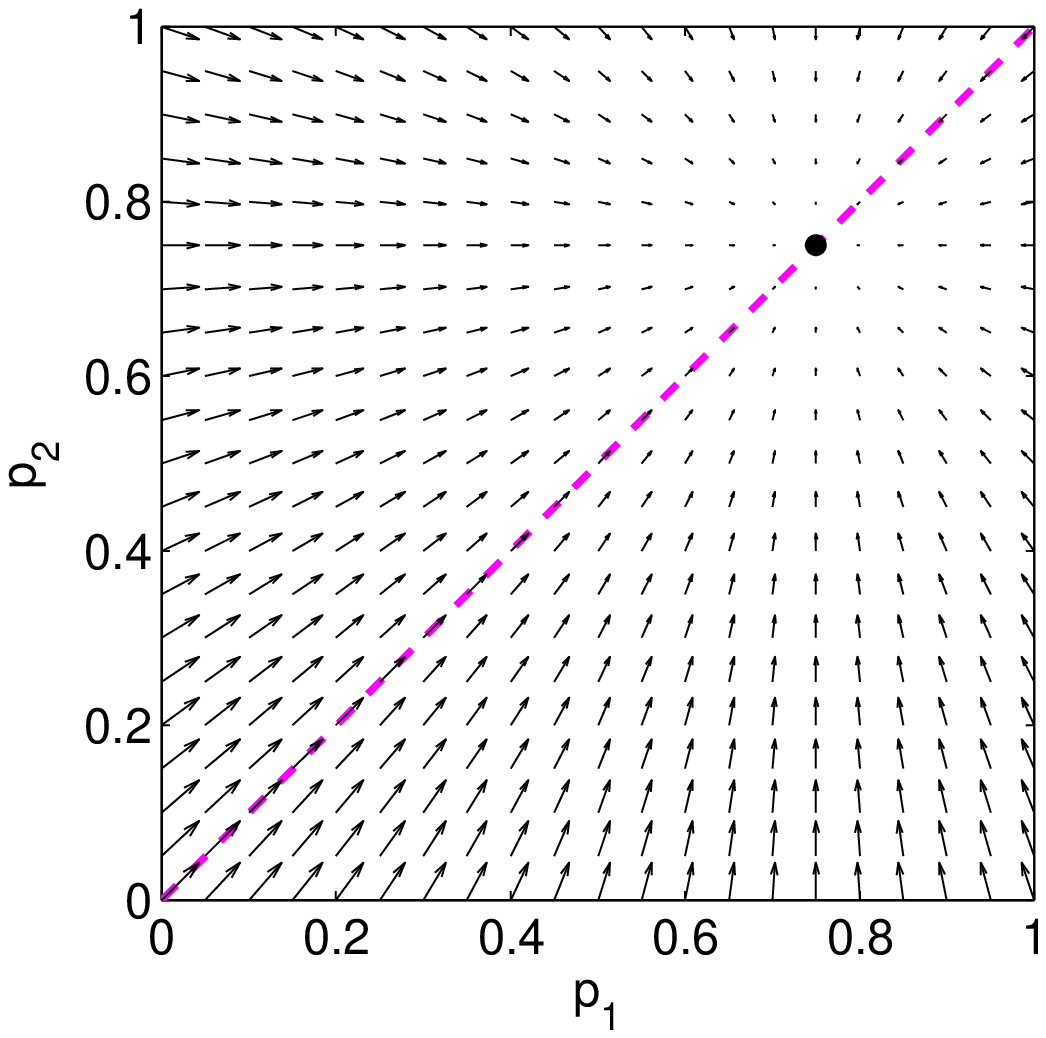}
\caption{Vector flow diagram representation of the co-action solution
of the 2-person Prisoners Dilemma game for temptation payoffs (left)
$T = 1.5$ and (right) $T = 3.5$. The abscissae and ordinate correspond
to the probabilities ($p_1$ and $p_2$) that players 1 and 2,
respectively, choose to cooperate. The broken line represents the
situation where the two agents have the same probability of
cooperation. The filled circles represent the unique co-action
equilibrium for each value of $T$ corresponding to the players
cooperating with equal probability ($=1$ for $T=1.5$ and $=0.75$ for
$T=3.5$).
}
\label{fig:3}       
\end{figure}

By contrast, using the co-action solution concept, for low values of
$T$ we observe mutual cooperation (i.e., $p_1 = p_2 = 1$) as the
stable equilibrium of the system (Fig.~\ref{fig:3}, left). For larger
values of $T$, the stable equilibrium corresponds to a mixed strategy, $0 <
p_1 = p_2 < 1$ (Fig.~\ref{fig:3}, right). Thus, as discussed in detail
in Ref.~\cite{upcoming}, using the co-action concept for solving PD 
we can show that selfish agents trying to maximize their individual
payoffs can also achieve the state of maximum collective benefit. This
resolves a contentious aspect associated with the Nash solution of PD,
where the agents end up worse off in trying to optimize their
individual payoffs~\cite{Morgan2009}.

\subsection{Chicken}
The Game of Chicken (also referred to as Snowdrift)~\cite{osborne} is another
well-studied 2-person game which is relevant in the
context of social interactions~\cite{rapoport} as well as
evolutionary biology~\cite{maynard} (where it is also known as
Hawk-Dove). The game represents a strategic interaction between two
players, who are driving towards each other in a potential collision
course, and have the choice between ``chickening out'', i.e., 
swerving away from the path of the other (Action~1) or continuing
on the path (Action~2). Thus, the choices correspond to being docile
or aggressive, respectively. 
If both players decide to swerve away, they receive the payoff $R$,
while if one swerves and the other continues on the path, the former
loses face (getting the payoff $S$) and the latter wins (payoff $T$).
However, the worst possible outcome corresponds to when both players
continue on the path, eventually resulting in a collision which is
associated with payoff $P$. The hierarchical relation between the
payoffs in Chicken is $T>R>S>P$, which suggests that a player will
benefit from being aggressive as long as the other is docile, but is
better off being docile if it is sure that the
other will play aggressively, as the cost of mutually aggressive
behavior is  high.

\begin{figure}[tbp]
\sidecaption[t]
\includegraphics[width=.45\linewidth]{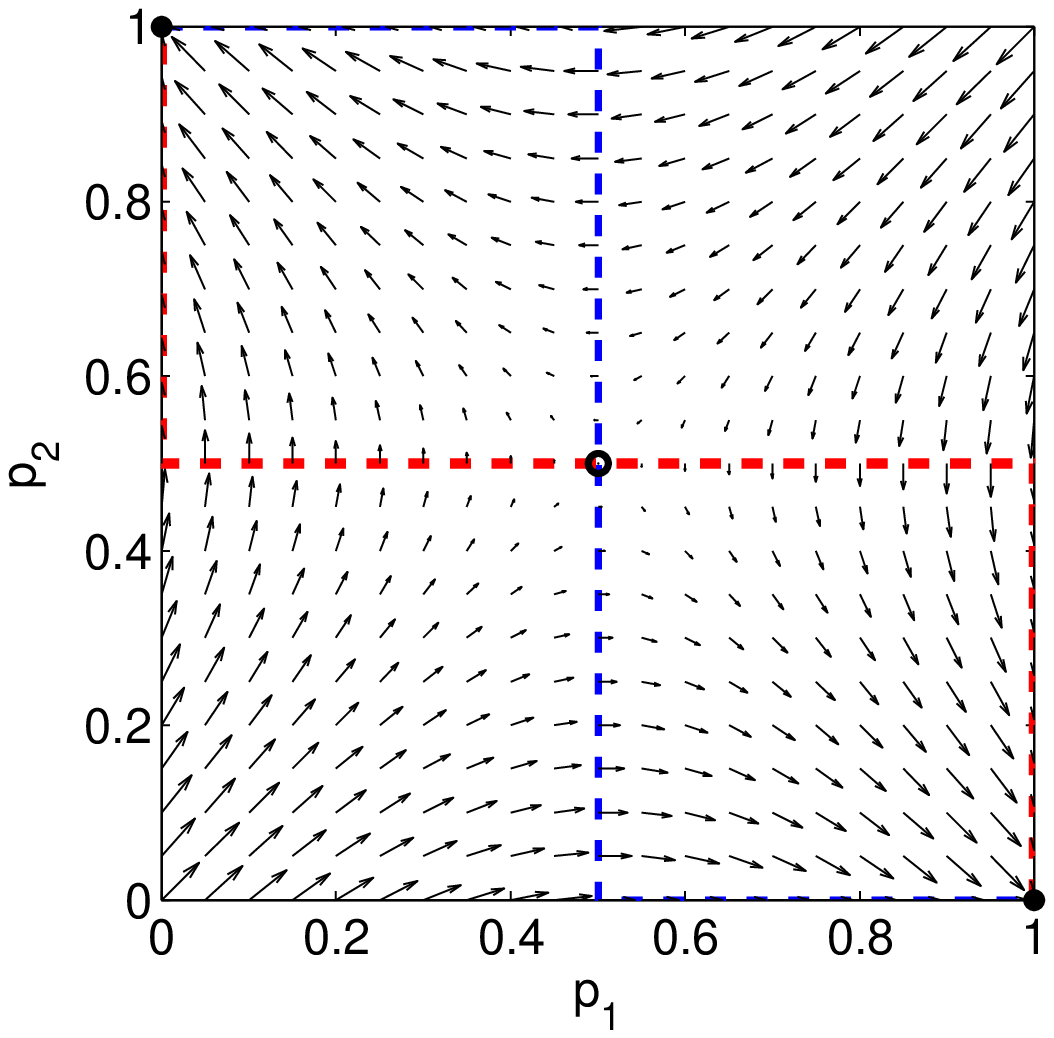}
\includegraphics[width=.45\linewidth]{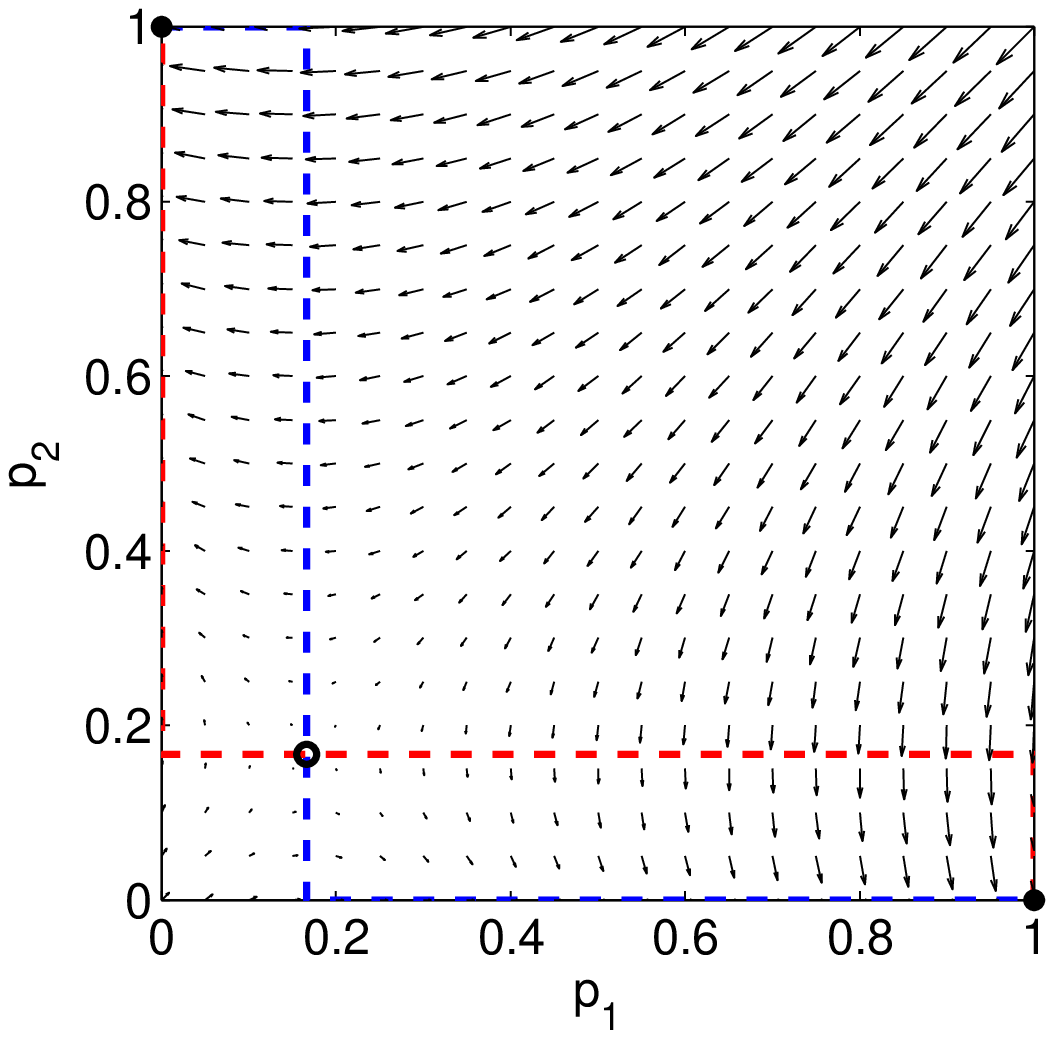}
\caption{Vector flow diagram representation of the Nash solution of the 2-person Chicken game for ``temptation'' payoffs (left) $T = 1.5$ and (right) $T = 3.5$. The abscissae and ordinate correspond to the probabilities ($p_1$ and $p_2$) that players 1 and 2, respectively, choose to be docile (i.e., non-aggressive). The broken lines represent the best response (or reaction) correspondence of the players (red for player 1, blue for player 2). The intersections of the lines, represented by unfilled and filled circles, represent the unstable and stable Nash equilibria respectively. The stable equilibria correspond to the pure strategy combination corresponding to one player being aggressive, the other being docile, while the unstable equilibrium in each case corresponds to a mixed strategy.
}
\label{fig:4}       
\end{figure}

Analyzing this game using the dynamical perspective described earlier
yields the flow diagram shown in Fig.\ref{fig:4} (obtained for two
different values of $T$, with $S = 0.5$) on using the Nash
solution concept. As can be seen, two of the multiple Nash equilibria
are stable, corresponding to the pure strategies (i) $p_1 =1, p_2 = 0$
and (ii) $p_1 = 0, p_2 = 1$ (i.e., when one player is aggressive, the
other is docile). The remaining equilibrium is an unstable mixed
strategy located on the $p_1 = p_2$ line (which defines the separatrix
demarcating the basins of attraction of the two stable equilibria).
With increasing $T$, the unstable
equilibrium - which dynamically corresponds to a saddle point in the
$p_1 - p_2$ plane - moves closer to $p_1 = 0, p_2 = 0$ corresponding to mutual
aggression.

\begin{figure}[tbp]
\sidecaption[t]
\includegraphics[width=.45\linewidth]{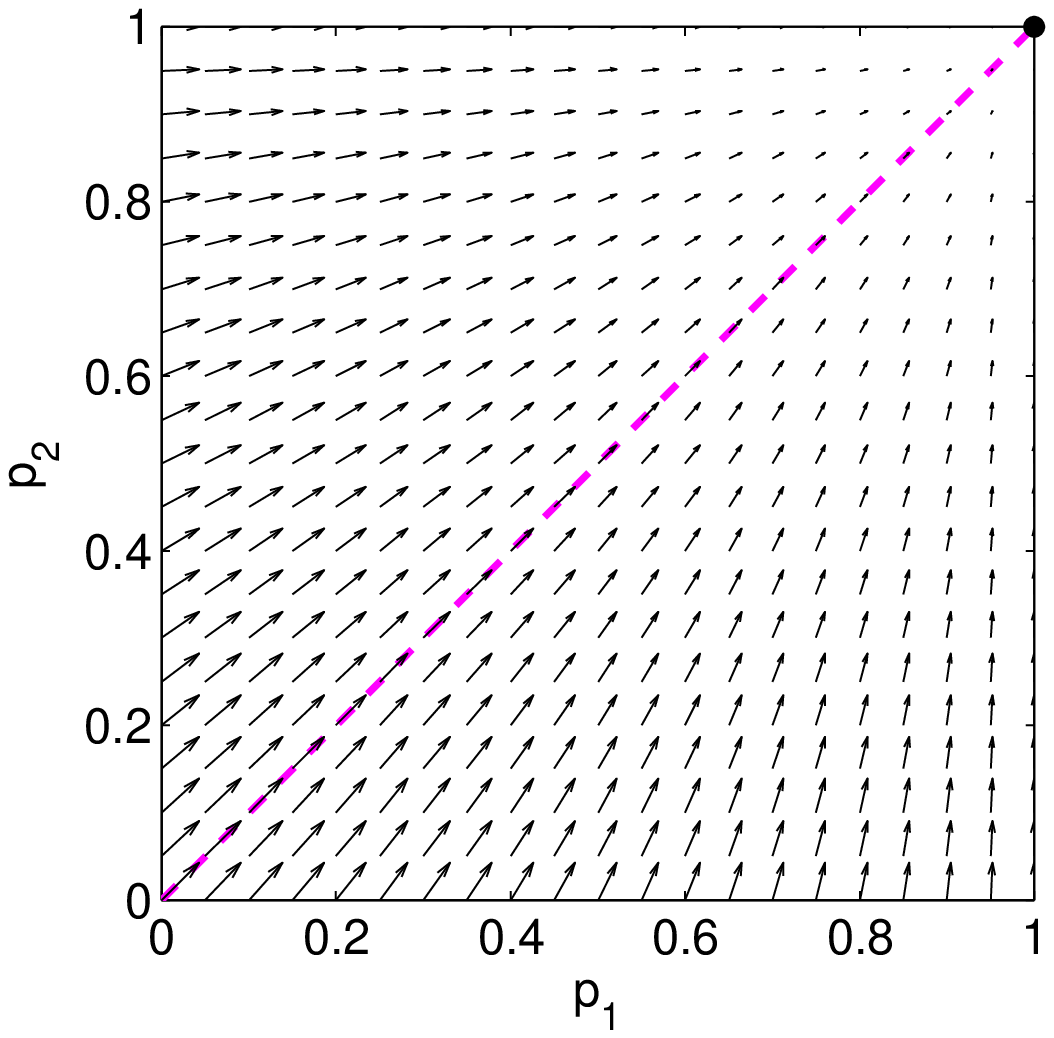}
\includegraphics[width=.45\linewidth]{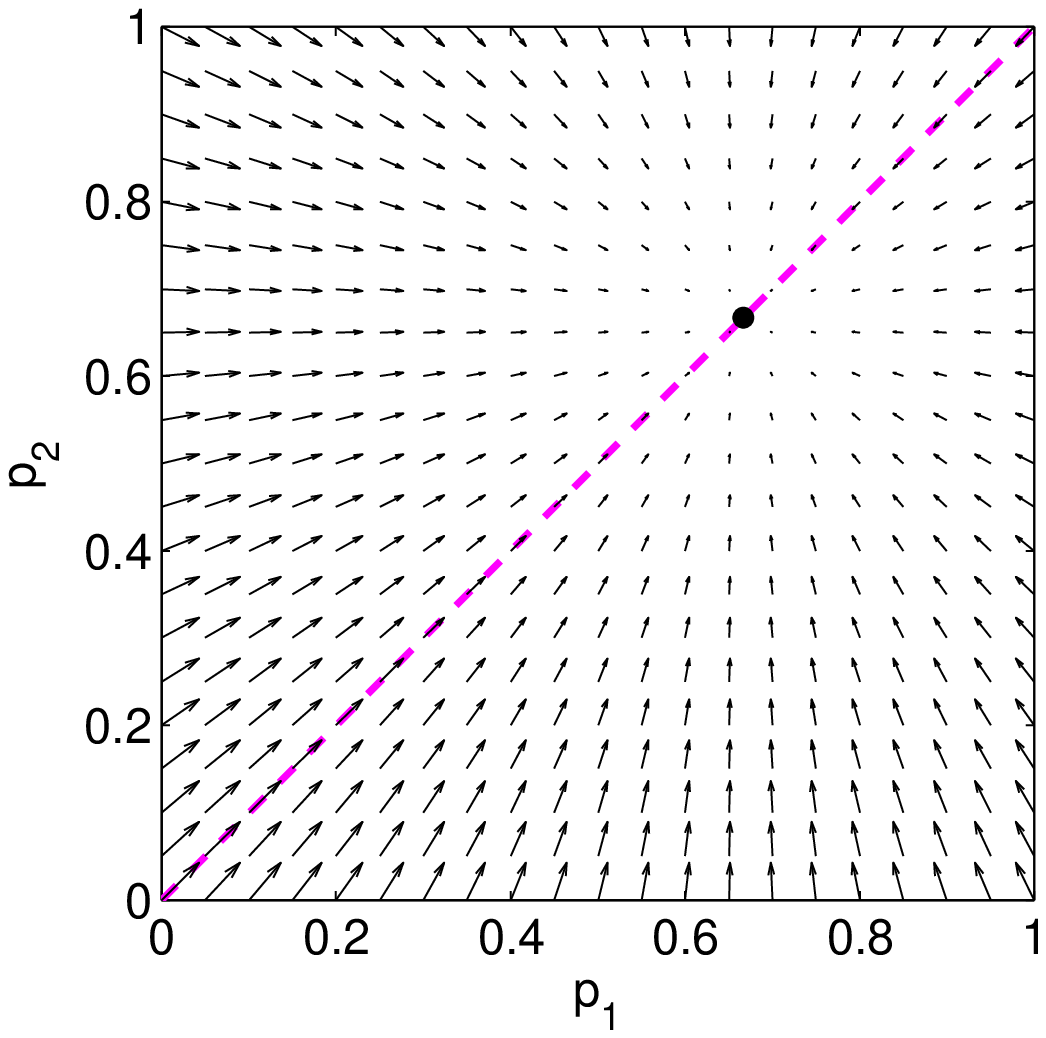}
\caption{Vector flow diagram representation of the co-action solution
of the 2-person Chicken game for ``temptation'' payoffs (left) $T =
1.5$ and (right) $T = 3.5$. The abscissae and ordinate correspond to
the probabilities ($p_1$ and $p_2$) that players 1 and 2,
respectively, choose to be docile (i.e., non-aggressive).  The broken
line represents the situation where the two agents have the same
probability of being docile. The filled circles represent the unique
co-action equilibrium for each value of $T$ corresponding to the
players choosing to be docile with equal probability ($=1$ for $T=1.5$
and $=2/3$ for $T=3.5$).
}
\label{fig:5}       
\end{figure}

Using the co-action solution concept gives rise to a qualitatively
different solution, as seen in the flow diagrams in Fig.~\ref{fig:5}. 
When $T$ is low, the system has a stable equilibrium at $p_1=1$,
$p_2=1$, i.e., both agents choose docile behavior to avoid the
potential damages associated with mutual aggression. For higher values
of $T$ the stable equilibrium is a mixed strategy $0 < p_1 = p_2 < 1$.
As in PD, the co-action paradigm yields a single, stable solution of
the game.

\subsection{Stag-Hunt}
The last of our examples, Stag-Hunt is a 2-person game that has been
studied in the context of emergence of coordination in social
interactions~\cite{skyrms}. The game represents a strategic
interaction between two players who have to choose between hunting
stag (Action~1) or hunting hare (Action~2). A hare may
be caught by a single agent but is worth less than a stag. On the
other hand, hunting a stag is successful only if both agents 
hunt for it. Thus, if both agents cooperate by hunting stag they
receive the highest payoff $R$. On the other hand, if they both choose
to hunt hare, they receive the payoff $P$. However, if one chooses to
hunt hare while the other goes for a stag, then the former receives
the payoff $T$ while the latter receives the worst possible payoff
$S$.
Thus, in Stag-Hunt, the hierarchical relation between the payoffs is
$R>T\geq P > S$, which suggests that while choosing to hunt hare may
be the safer option, there is a possibility of doing much better by
choosing to hunt stag if one is confident that the other
will also do the same.
\begin{figure}[tbp]
\sidecaption[t]
\includegraphics[width=.45\linewidth]{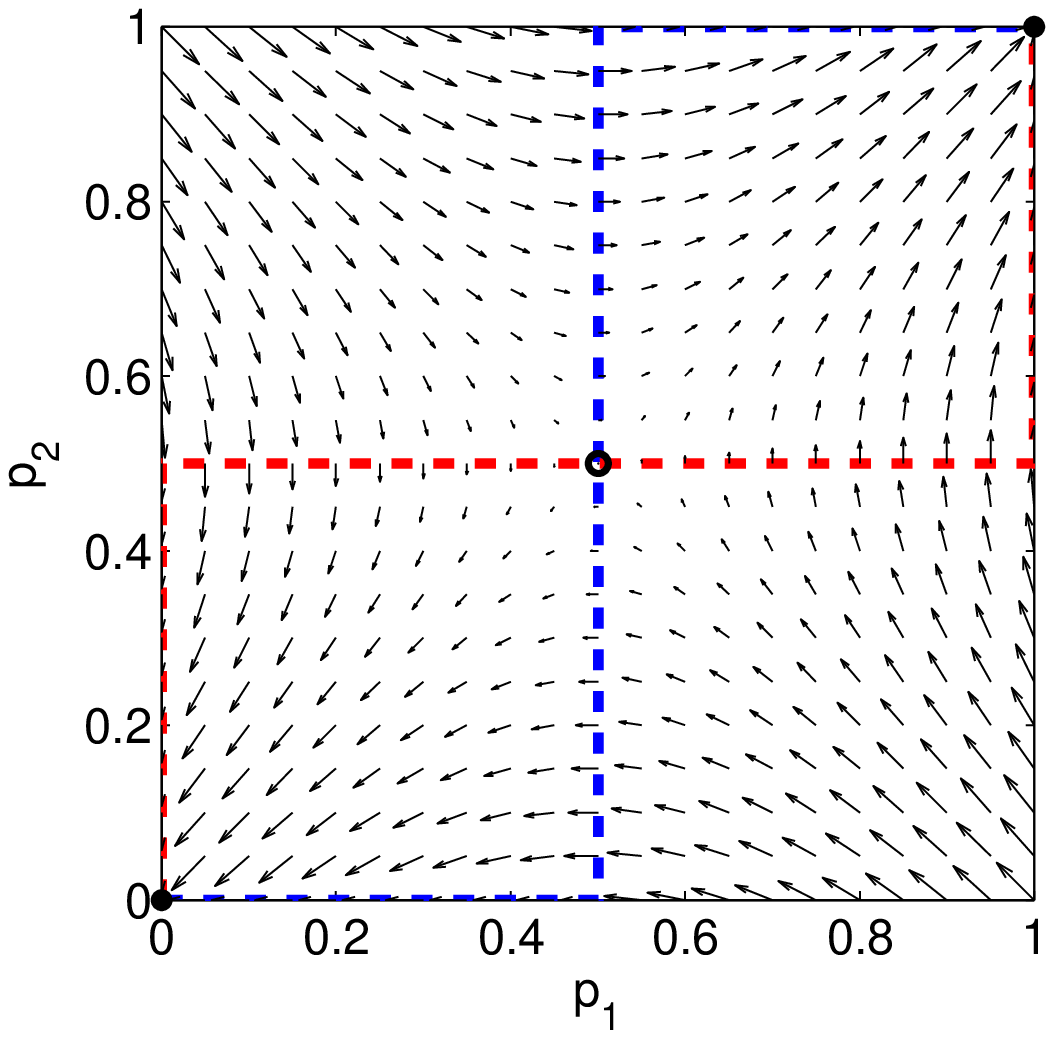}
\includegraphics[width=.45\linewidth]{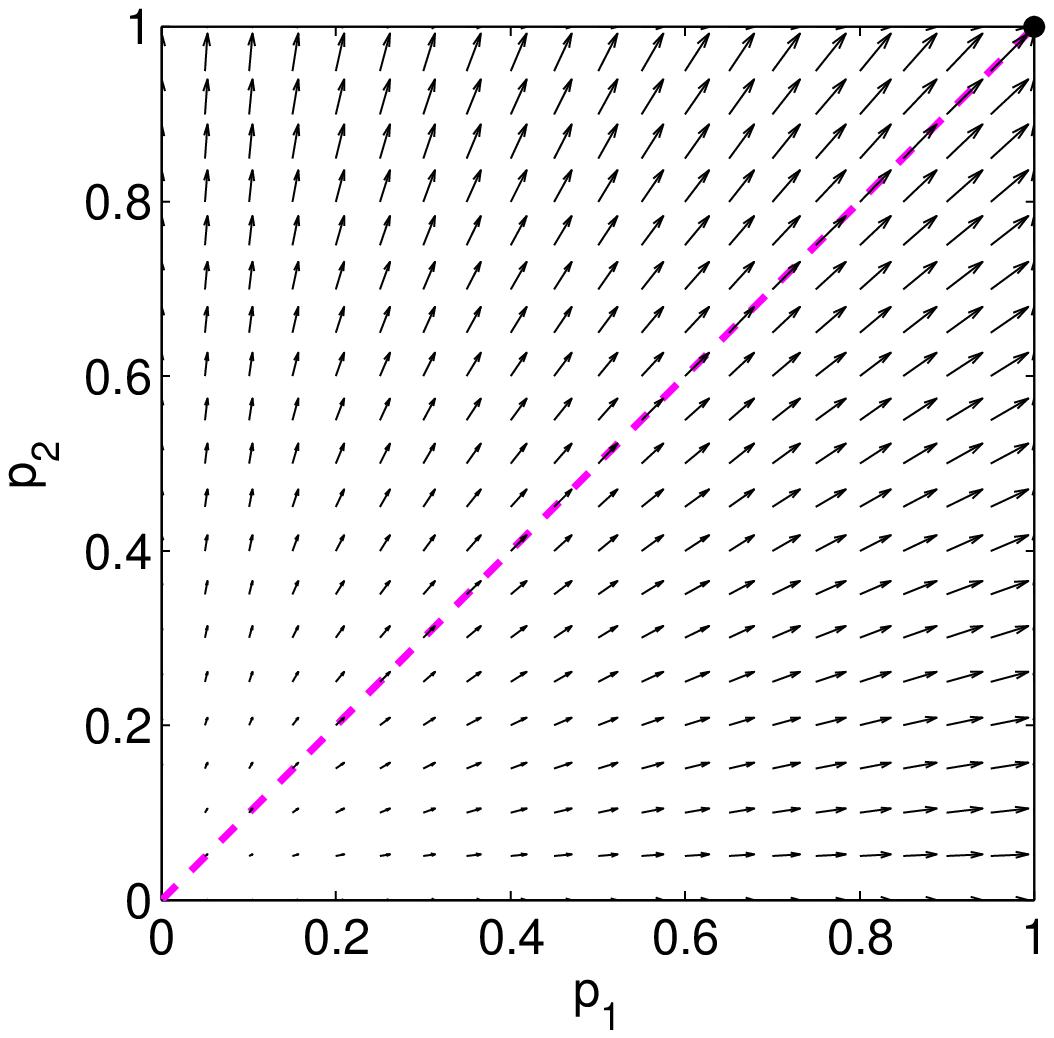}
\caption{Vector flow diagram representation of the (left) Nash and (right) co-action solutions of the 2-person Stag-Hunt game for $T= 0.5$ and $S = -0.5$. The abscissae and ordinate correspond to the probabilities ($p_1$ and $p_2$) that players 1 and 2, respectively, choose `Stag' instead of `Hare'. (left) The broken lines represent the best response (or reaction) correspondence of the players (red for player 1, blue for player 2). The broken line represents the situation where the two agents have the same probability of choosing 'stag'. }
\label{fig:6}       
\end{figure}

The vector flow diagrams for Nash and co-action solution concepts in the Stag-Hunt are shown in
Fig.\ref{fig:6} (obtained for $T = 0.5$ and $S = -0.5$). For Nash, as in the game of Chicken, there are
three equilibria (Fig.~\ref{fig:6},~left), of which the pure strategies, corresponding to (i) $p_1 =1, p_2 = 1$
and (ii) $p_1 = 0, p_2 = 0$ are stable (i.e., when both players hunt for stag or when both players hunt hare). The remaining equilibrium is an unstable mixed strategy located on the $p_1 = p_2$ line which again defines the separatrix
demarcating the basins of attraction of the two stable equilibria.

The co-action solution for the games (Fig.~\ref{fig:6},~right) is a simple one in which both agents always choose hunting stag. i.e, $p_1 = p_2 = 1$. Thus, under the co-action concept, the players always converge to the best possible outcome. In this case, there is no mixed strategy equilibrium for any value of the parameters.

%
%

\section{Conclusions}
\label{sec:4}
In this article we have shown that using a dynamical perspective
allows us a visually appealing way to differentiate between two
solution concepts, viz., Nash and co-action, for 2-person, symmetric games which lead to
spectacularly different conclusions. To illustrate these differences
in details we looked at three examples of such games in detail:
Prisoners Dilemma, Chicken and Stag-Hunt. In all of these games, one
action - in particular, Action~1 in the terminology used here -
corresponds to the players being \textquotedblleft nicer\textquotedblright\; to each other (e.g., cooperating
in PD, etc.) compared to the other action. The vector flow diagrams
generated by the approach presented here clearly show that co-action
more often results in nicer strategies being converged at by the
agents than in the case for Nash.
Our results are intriguing in view of the experimental literature on
2-person games (see discussion in Ref.~\cite{Morgan2009}), in particular PD, which seems to suggest that when
these games are played between real human individuals they tend to be
far nicer than suggested by the Nash solution. 

\begin{acknowledgement}
We thank Deepak Dhar for useful discussions and Shakti N Menon for a careful reading of the manuscript.
\end{acknowledgement}

\end{document}